\let\ifarxiv=\iftrue     
\pdfoutput=1

\ifarxiv

\documentclass[12pt,a4paper]{article}
\usepackage[a4paper,text={450pt,650pt},centering]{geometry}

\fi

\ifarxiv\else

\documentclass[11pt,a4paper]{article}
\usepackage{mathptmx}
\usepackage[a4paper,text={130mm,198mm}]{geometry}

\fi

\usepackage{amsmath,amssymb}
\usepackage[bookmarks=true,hyperfigures=true]{hyperref}
\usepackage{graphicx}
\usepackage[nosort]{cite}
\usepackage[bulletsep]{collref}

\let\oldbfseries=\bfseries
\let\oldmdseries=\mdseries
\let\oldnormalfont=\normalfont
\renewcommand{\bfseries}{\oldbfseries\boldmath}
\renewcommand{\mdseries}{\oldmdseries\unboldmath}
\renewcommand{\normalfont}{\oldnormalfont\unboldmath}

\allowdisplaybreaks[3]

\numberwithin{equation}{section}

\usepackage[font=small,labelfont=bf,width=0.85\textwidth]{caption}

\providecommand{\hypersetup}[1]{}

\hypersetup{plainpages=false}
\hypersetup{pdfpagemode=UseNone}
\hypersetup{bookmarksnumbered=true}
\hypersetup{pdfstartview=FitH}
\hypersetup{colorlinks=false}
\hypersetup{citebordercolor={.5 1 .5}}
\hypersetup{urlbordercolor={.5 1 1}}
\hypersetup{linkbordercolor={1 .7 .7}}

\providecommand{\href}[2]{#2}
\providecommand{\arxivlink}[1]{\href{http://arxiv.org/abs/#1}{arxiv:#1}}

\newcommand{\nn}{\nonumber}

\newcommand{\earel}[1]{\mathrel{}&\hspace{-2\arraycolsep}#1\hspace{-2\arraycolsep}&\mathrel{}}
\newcommand{\eq}{\earel{=}}

\def\[{\begin{equation}}
\def\]{\end{equation}}
\def\<{\begin{eqnarray}}
\def\>{\end{eqnarray}}

\newcommand{\sfrac}[2]{{\textstyle\frac{#1}{#2}}}
\newcommand{\half}{\sfrac{1}{2}}

\newcommand{\quarter}{\sfrac{1}{4}}
\newcommand{\ind}{\varrho}

\def \t {\tilde}

\def \bp {\begin{pmatrix}}  \def \epm {\end{pmatrix}}

\def \la {\label}
\def \l {\lambda}
\def \sql {{\sqrt \l}}
\def \adss {$AdS_5 \times S^5$ }
\newcommand{\rf}[1]{(\ref{#1})}

\def \J {{\rm J}}
\def \S {{\rm S}} 
\def \E {{\rm E}}
\def \F {G}
\def \T {{T}}

\newcommand{\alg}[1]{\mathfrak{#1}}
\newcommand{\grp}[1]{\mathrm{#1}}

\newcommand{\grSU}{\grp{SU}}

\newcommand{\grPSU}{\grp{PSU}}

\newcommand{\unit}{\mathbb{I}}

\newcommand{\order}{\mathcal{O}}

\newcommand{\comma}{\quad,\quad}

\newcommand{\Bsi}{\Upsilon}

\newcommand{\lAA}{{a}}
\newcommand{\rAA}{{\dot{a}}}
\newcommand{\laa}{{\alpha}}
\newcommand{\raa}{{\dot{\alpha}}}
\newcommand{\lBB}{{b}}

\newcommand{\lbb}{{\beta}}

\newcommand{\lCC}{{c}}

\newcommand{\lcc}{{\gamma}}

\newcommand{\lDD}{{d}}

\newcommand{\ldd}{{\delta}}

\newcommand{\cpp}{\energy' p - \energy p'}

\newcommand{\Smatrix}{\mathbb{S}}  
\newcommand{\smatrix}{\mathbf{S}}   
\newcommand{\Tmatrix}{\mathbb{T}}  
\newcommand{\tmatrix}{\mathrm{T}}   

\newcommand{\Atmatrix}{\mathrm{A}}
\newcommand{\Btmatrix}{\mathrm{B}}
\newcommand{\Ctmatrix}{\mathrm{C}}
\newcommand{\Dtmatrix}{\mathrm{D}}
\newcommand{\Etmatrix}{\mathrm{E}}
\newcommand{\Ftmatrix}{\mathrm{F}}
\newcommand{\Gtmatrix}{\mathrm{G}}
\newcommand{\Htmatrix}{\mathrm{H}}
\newcommand{\Ktmatrix}{\mathrm{K}}
\newcommand{\Ltmatrix}{\mathrm{L}}

\newcommand{\levi}{\epsilon}
\newcommand{\energy}{\varepsilon}

\newcommand{\lrbrk}[1]{\left(#1\right)}

\newcommand{\biggsbrk}[1]{\biggl[#1\biggr]}

\newcommand{\ket}[1]{\mathopen{|}#1\mathclose{\rangle}}

\makeatletter
\def\mr@ignsp#1 {\ifx\:#1\@empty\else #1\expandafter\mr@ignsp\fi}%
\newcommand{\multiref}[1]{\begingroup
\xdef\mr@no@sparg{\expandafter\mr@ignsp#1 \: }%
\def\mr@comma{}%
\@for\mr@refs:=\mr@no@sparg\do{\mr@comma\def\mr@comma{,}\ref{\mr@refs}}%
\endgroup}
\makeatother

\newcommand{\hypref}[2]{\ifx\href\asklfhas #2\else\href{#1}{#2}\fi}

\newcommand{\secref}[1]{Sec.~\multiref{#1}}

\newcommand{\figref}[1]{Fig.~\multiref{#1}}
\renewcommand{\eqref}[1]{(\multiref{#1})}

\makeatletter
\newlength{\apb@width}
\newcommand{\autoparbox}[2][c]{\settowidth{\apb@width}{#2}\parbox[#1]{\apb@width}{#2}}

\makeatother

\begin{document}


\thispagestyle{empty}
\phantomsection
\addcontentsline{toc}{section}{Title}

\begin{flushright}\footnotesize%
\texttt{AEI-2010-177},
\texttt{\arxivlink{1012.3987}}\\
overview article: \texttt{\arxivlink{1012.3982}}%
\vspace{1em}%
\end{flushright}

\begingroup\parindent0pt
\begingroup\bfseries\ifarxiv\Large\else\LARGE\fi
\hypersetup{pdftitle={Review of AdS/CFT Integrability, Chapter II.2: Quantum Strings in AdS5xS5}}%
Review of AdS/CFT Integrability, Chapter II.2:\\
Quantum Strings in AdS$_5\times$S$^5$
\par\endgroup
\vspace{1.5em}
\begingroup\ifarxiv\scshape\else\large\fi%
\hypersetup{pdfauthor={Tristan McLoughlin}}%
Tristan McLoughlin
\par\endgroup
\vspace{1em}
\begingroup\itshape
Max-Planck-Institut f\"ur Gravitationsphysik,\\
 Albert-Einstein-Institut, \phantom{$^\ddag$}\\
Am M\"uhlenberg 1, D-14476 Potsdam, Germany
\par\endgroup
\vspace{1em}
\begingroup\ttfamily
tmclough@aei.mpg.de
\par\endgroup
\vspace{1.0em}
\endgroup

\begin{center}
\includegraphics[width=5cm]{TitleII2.mps}
\vspace{1.0em}
\end{center}

\paragraph{Abstract:}
We review the semiclassical analysis of strings in AdS$_5\times$S$^5$
with a focus on the relationship to the underlying integrable structures. 
We discuss the perturbative calculation of energies for strings with large charges,
using the  folded string spinning in AdS$_3$ $\subset$ AdS$_5$ as our main example. 
Furthermore, we review the perturbative light-cone quantization of the string theory
and the calculation of the worldsheet S-matrix.

\ifarxiv\else
\fi
\hypersetup{pdfsubject={MSC (2010): 81T30, 81U15, 83E30}}%

\ifarxiv\else
\fi
\hypersetup{pdfkeywords={Strings in AdS space, quantum corrections, integrability.}}%

\newpage


\section{Introduction}
\label{sec:intro}

The semiclassical  study of  strings in AdS$_5\times$S$^5$
 has played a key 
role in extending our understanding of the AdS/CFT correspondence beyond the supergravity
approximation. 
The analysis of quantum corrections to the energies of strings with large 
charges has gone hand-in-hand with the discovery and application of the integrable
structures present in the  duality. 
In particular, it has been important for comparison with the Bethe
ansatz predictions for the anomalous dimensions of long operators and to understand 
the finite size corrections of short operators. 

Due to the presence of Ramond-Ramond fields one must
make use of the Green-Schwarz formalism for the string action, adapted to the AdS$_5\times$S$^5$
geometry \cite{Metsaev:1998it} (see \cite{chapSpinning} for a brief introduction),
\footnote{One can also study strings in different backgrounds, AdS$_4\times$$\mathbb{C}$P$^3$
is of particular interest where many results parallel the AdS$_5\times$S$^5$ case. 
See \cite{chapN6}.}
 which to quadratic order in fermionic fields is
\<
\label{eq:action}
I=-\frac{\sqrt{\lambda}}{4\pi}\int d^2\sigma h^{ab}\, G_{\mu\nu}\partial_a x^\mu\partial_b x^\nu-
i\frac{\sqrt{\lambda}}{2\pi}\int d^2\sigma \left( h^{ab}\delta^{IJ}-\epsilon^{ab} s^{IJ}\right)  {\bar \theta}^I\ind_a D_b\theta^J.
\>
Here we have used the rescaled worldsheet metric $h^{ab}=\sqrt{-g}g^{ab}$,  
the induced Dirac matrices 
$\ind_a=\partial_a x^\mu E_\mu{}^A\Gamma_A$
and the covariant derivative
\<
D_a\theta^I=\left(\partial_a +\frac{1}{4} \partial_a x^\mu \omega_{\mu}{}^{ AB}\Gamma_{AB}\right)\theta^{I}
+\frac{1}{2}\ind_a \Gamma_{01234} \epsilon^{IJ}\theta^J~.
\>
Directly quantizing this action is beyond current methods and one must 
take a perturbative approach, expanding about a given classical solution in powers 
of the effective string tension, $\sqrt{\lambda}$.
A classical solution 
is characterised by the conserved charges corresponding to the AdS energy, $E$, two AdS
spins, $S_i$,  and three angular momenta
 of the sphere, $J_s$, in addition to any parameters specifying further properties
 of the string  such as non-trivial winding. 
The Virasoro conditions provide a constraint on these parameters and for
the solutions we are interested in 
we can express the string energy as a function of the
 remaining charges: $E=E({S}_i, {J}_s;k_r)$.
In the 
semiclassical approach one takes a string solution where one or more
of the rescaled charges are finite, ${\S}_i=\tfrac{S_i}{\sqrt{\lambda}}$ 
or ${\J}_s= \tfrac{J_s}{\sqrt{\lambda}}$,
and computes the worldsheet loop corrections to the energy as an expansion in large tension,
\<
E=\sqrt{\lambda}\Big[{\E}_0({\rm S}_i, {\J}_s;k_r)+\frac{1}{\sqrt{\lambda}}{\E}_1({\S}_i, {\J}_s;k_r)+\frac{1}{\lambda}{\E}_2({\S}_i, {\J}_s;k_r)+\dots \Big]~.
\>
In general, calculating these corrections involves gauge-fixing the diffeomorphism and kappa
gauge invariance, 
and studying the fluctuations of the fields -- bosonic, fermionic 
and conformal ghosts from gauge fixing -- about the classical solution. 
An important point is that all UV divergences of the worldsheet theory 
 cancel and, relatedly, the conformal anomaly vanishes once the contribution from
the path integral measure is accounted for;  thus the semiclassical expansion is well 
defined. On general grounds this is expected as the string theory 
is of critical dimension and it was explicitly shown at one-loop in \cite{Drukker:2000ep} 
and \cite{Frolov:2002av}.
\footnote{ Particular care must be taken with the fermionic fields. Importantly, they couple to the worldsheet metric rather than the zweibein and so
contribute to the conformal anomaly four times the usual 2-d Majorana fermion amount. 
%
}
A solution which has
played a particularly important role in 
our quantitative understanding of the AdS/CFT duality is the spinning folded string in AdS$_5$, 
introduced in \cite{Gubser:2002tv} and whose semiclassical 
analysis was initiated in \cite{Frolov:2002av}.
In the large spin limit \cite{Gubser:2002tv, Belitsky:2006en, Frolov:2006qe}, the difference between its energy $E$ and spin $S$
scales like $\ln S$  with the 
coefficient being the universal scaling function, $f(\lambda)$.  This function
provided the first example of a result interpolating 
between weak and strong coupling which can be calculated from the all-order asymptotic Bethe
ansatz (ABA) \cite{Eden:2006rx, Beisert:2006ez} (see \cite{chapABA, chapSMat, chapSProp}
for a review of the all-order ABA). The one and two-loop semiclassical calculations 
\cite{Frolov:2002av, Roiban:2007dq, Roiban:2007jf, Giombi:2009gd} 
have been shown 
to match 
the predictions of the string ABA \cite{Arutyunov:2004vx, Staudacher:2004tk, Beisert:2005fw} 
using the one-loop phase factor \cite{Beisert:2005cw, Hernandez:2006tk, Freyhult:2006vr} and its all-order 
generalisation \cite{Beisert:2006ib, Beisert:2006ez} 
in a very non-trivial test of the duality and its quantum integrability (see \cite{chapTwist}
for a review of the ABA calculation and references). We will discuss this 
solution, its generalisations and related solutions in \secref{sec:sfs}.
While for the
most part we focus on closed strings, similar  semiclassical analysis  has also been applied to open
strings:  duals to  cuspy Wilson loops,  to
Wilson loops describing  ``quark--anti-quark" systems, \cite{Drukker:2000ep, Forste:1999qn, Sonnenschein:1999yb, Kinar:1999xu,Forini:2010ek},
 to Wilson loops describing high energy scattering
\cite{Janik:2000pp, PandoZayas:2003yb} and 
more recently, dimensionally reduced amplitudes \cite{Kruczenski:2007cy}.

Another solution which has played a crucial
role in our understanding of the quantum string in AdS$_5\times$S$^5$ is the BMN string, 
\cite{Berenstein:2002jq} \cite{Gubser:2002tv} see also
\cite{chapSpinning},
which is the BPS solution dual to the ferromagnetic vacuum of the spin chain description 
of the gauge theory. This solution is the natural vacuum state in the light-cone quantization 
of the worldsheet theory where the physical Hamiltonian, $H_{l.c.}$, is 
proportional to $P_-=E-J$,
with $J$  one of the sphere angular momenta. 
\footnote{There are essentially two ways to fix the light-cone
gauge in AdS$_5\times$S$^5$, which differ by picking inequivalent
light-cone geodesics. In one case, which is possible only in the
Poincar\'e patch, the light-cone directions lie entirely in
AdS$_5$ \cite{Metsaev:2000yf, Metsaev:2000yu}.
In our case the light-cone is shared
between
AdS$_5$ and  S$^5$ e.g.
\cite{Callan:2003xr, Callan:2004uv, Arutyunov:2005hd, Frolov:2006cc, Arutyunov:2006gs}.}
Finding quantum string energies, $E$, corresponds to computing 
the spectrum of the $H_{l.c.}$.
 Unfortunately the exact light-cone Hamiltonian
has a non-polynomial form \cite{Callan:2004uv, Arutyunov:2004yx} and is not a suitable starting point
for ``first-principles" quantization. One can, however, solve for the spectrum
perturbatively. At leading order the theory is simply that of free massive fields
\cite{Berenstein:2002jq, Metsaev:2001bj, Metsaev:2002re}
while at subleading orders \cite{Parnachev:2002kk,
Callan:2003xr, Callan:2004uv, Callan:2004ev, McLoughlin:2004dh, Frolov:2006cc} the 
interactions are somewhat more complicated and, due to the gauge fixing, do not 
respect worldsheet Lorentz invariance. 
Alternatively, as the worldsheet theory is integrable,
it is possible to find the spectrum of the decompactified theory, via the ABA,  by calculating the
worldsheet S-matrix \cite{Staudacher:2004tk}, \cite{Arutyunov:2004vx, Beisert:2005fw}. 
A review of the exact form of this S-matrix and its properties
can be found \cite{chapSMat, chapSProp}, in this review we will restrict ourselves 
to briefly describing its perturbative calculation (for a more thorough review
see \cite{Arutyunov:2009ga}).

\section{Quantum spinning strings}
We will, as an illustrative example, consider the 
the folded spinning string \cite{Gubser:2002tv}, \cite{Frolov:2002av}, 
see also \cite{chapSpinning}. 
\begin{figure}\centering
\begin{tabular}{cc}
\includegraphics[scale=0.25]{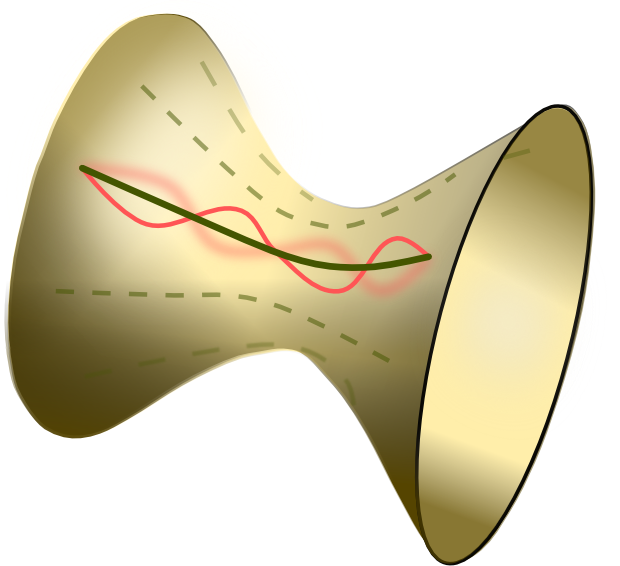}
\hspace{2cm}&
\includegraphics[scale=0.33]{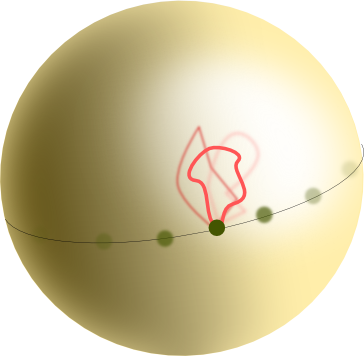}\\
~ & ~ \\
 (a)\hspace{2cm}&  (b)
 \end{tabular}
\caption{In (a) we show the classical folded spinning string moving in AdS$_3$ $\subset$ AdS$_5$
at a certain time (dark solid line) and earlier/later times (dashed lines). The quantum fluctuations, corresponding to oscillations transverse (light wavy lines) to the classical solution, acquire mass due to the background curvature. In (b) we show the motion of the string on the sphere, essentially a point moving along a great circle, with its fluctuations again seeing more of the geometry.   } \label{fig:Quantum_spinning_string}
\end{figure}
This solution describes a string  extended and rotating with spin, $S$,  in an AdS$_3$  subspace
of AdS$_5$ while additionally moving
along a great circle of the S$^5$ with angular momentum $J$ (see \figref{fig:Quantum_spinning_string}). 
In terms of the global coordinates 
\<
\label{eq:global_coor}
ds_{{\rm AdS}_5}^2\eq-\cosh^2 \rho\ dt^2+d\rho^2+\sinh^2\rho \left( d\theta^2+\cos^2\theta\ d\phi_1^2+\sin^2\theta\ d\phi_2^2\right)~,\\
ds_{{\rm S}^5}^2\eq+\cos^2 \gamma\ d\varphi_3^2+d\gamma^2+\sin^2\gamma \left( d\psi^2+\cos^2\psi\ d\varphi_1^2+\sin^2\psi\ d\varphi_2^2\right)~,
\>
the string solution is given by $\theta=\gamma=\psi=\tfrac{\pi}{2}$,
\<
\label{eq:gensol}
t=\kappa \tau~, ~~ \phi_2=\omega \tau~, ~~
\rho=\rho(\sigma)=\rho(\sigma+2\pi)~,~~
 \varphi_2=\nu\tau~.
\>
The equations of motion and the conformal constraints are satisfied provided 
\<
\rho''=(\kappa^2-\omega^2)\sinh\rho\cosh\rho~,~~~
~~~{\rho'}{}^2=\kappa^2\cosh^2\rho-\omega^2\sinh^2\rho-\nu^2~,
\>
and the other fields are zero.
This string can be thought of as four segments: the first, for $0\leq \sigma\leq \tfrac{\pi}{2}$, extends
from the origin of the AdS$_5$ space
along the radial direction to a maximum $\rho(\tfrac{\pi}{2})=\rho_0$  i.e. $\rho'(\tfrac{\pi}{2})=0$.
The string then turns and runs back along itself to the origin, this then repeats before the string closes
on itself.
In fact, this solution is generically rather complicated
however, 
in various limits it simplifies dramatically. 

\subsection{Quantum corrections}
\label{sec:qc}
It is possible to extract the one-loop correction to the energy by various means 
though, of course, all give identical results. 
The most direct method is to 
fix a physical gauge, such as light-cone, 
solve the resulting constraints and quantise the remaining 
degrees of freedom; the correction to the AdS energy of the string
is the correction to the two-dimensional energy of the vacuum state. 
However, for many purposes, and particularly for more complicated
solutions at higher orders, the most convenient method, 
introduced in this context by \cite{Roiban:2007jf, Kruczenski:2007cy, Roiban:2007dq} 
and most completely described in \cite{Roiban:2007ju, Giombi:2010fa}, is to relate the 
correction to the energy to the calculation of the worldsheet effective action.
\footnote{There is yet another method, essentially a generalisation of the WKB
formula, for finding the
leading quantum correction to periodic solutions
due to Daschen, Hasslacher and Neveu \cite{Dashen:1975hd}.
%
Such methods were applied to
the semiclassical quantization of the giant magnon \cite{Hofman:2006xt}
in   \cite{Minahan:2007gf, Papathanasiou:2007gd, Chen:2007vs} }
As in standard QFT, and in analogy with the 
thermodynamic Gibbs free energy, in the presence of 
a non-trivial background solution, $\varphi_{c}(x)$, the expectation value of the 
conjugate  source, $J(x)$, 
is given by the functional derivative of the effective action, $\Gamma[\varphi_c(x)]$, which is simply 
the Legendre transform of the vacuum energy functional. 
For the theory we are interested
in the sources are simply the conserved charge densities, such as $\E$, $\S$ and $\J$.
These are conjugate to time derivatives of the fields and 
so the background is specified by the constant parameters e.g. $\kappa$, $\omega$, and $\nu$. 
Thus
\<
\label{eq:eff_ac}
\frac{1}{ \T}\Gamma(\kappa, \omega,\nu )=-\frac{i}{ \T} \ln\ \langle e^{i H_{2d} \T}\rangle+  \kappa \langle  E \rangle -
{\omega} \langle S\rangle-{\nu} \langle J \rangle
\>
where $\T\rightarrow \infty$ is the worldsheet time interval. 
Due to the classical Virasoro constraints not all parameters are independent
 e.g. $\kappa=\kappa(\omega, \nu)$.
Furthermore,  the energy functional vanishes as $\langle H_{2d}\rangle=0$
due to the quantum conformal constraint.
The charges are thus found from the effective action by e.g. 
\<
\label{eq:qcharges}
\frac{1}{\T}\frac{\partial\Gamma(\omega, \nu)}{\partial \nu}=
\frac{\partial\kappa(\omega, \nu)}{\partial \nu}\langle E\rangle- \langle J \rangle~.
\>
Hence, we need only calculate the 
worldsheet effective action to determine the corrections to the string charges. 
In general, the leading quantum correction to  the effective action, $\Gamma_1$, is 
 found by expanding the Lagrangian, $L$, about a classical solution, 
 $\varphi=\varphi_c+\t \varphi$, and performing the Gaussian integral
\<
 \Gamma_1=\frac{i}{2}\log\ {\rm det}\Big[-\frac{\delta^2 L}{\delta\t\varphi\delta\t\varphi}\Big]
 =\frac{i}{2}{\rm Tr}\ \log\Big[-\frac{\delta^2 L}{\delta\t\varphi\delta\t\varphi}\Big]~.
 \>
For the string theory we must include not only the bosonic fluctuations
but also those of the fermionic  and the ghost fields which give inverses of determinants.

In general the  effective action is an extrinsic quantity. 
\footnote{Strictly speaking all our considerations are only valid in the large volume limit
and under the assumption that interactions are local.} This can be seen by
considering the simple case where the  quadratic fluctuation operator is given 
by $K=-\partial^2+m^2$ with constant masses, $m$. Fourier transformed this is
$\t K=-\omega^2+n^2+m^2$, and so 
\<
\Gamma_1=\frac{i \T}{2} \int \frac{d\omega}{2\pi}\sum_n \log\left(-\omega^2+n^2+m^2\right)
		=\frac{l\T}{2}\int \frac{ d^2p_E}{(2\pi)^2}\log \left(p_E^2+m^2\right)
\>
where in the last identity we have Wick 
rotated to Euclidean signature and taken the extent of the spatial direction, $l$, to also be large.
Note that by
performing the integration over $\omega$ in this constant mass case, or in fact for any 
stationary solution, one recovers the sum over fluctuation 
frequencies which gives the more common expression for the correction to the string energy
 c.f. appendix A \cite{Frolov:2002av}.
 \footnote{It is also possible to make use of the integrable structure and extract the
 fluctuation frequencies from the string algebraic curve. While this  powerful 
 method is widely used in the calculation of quantum corrections we will not
 discuss it here, but simply refer the reader to \cite{chapCurve} for a review and references.}

\subsection{Point-like BMN string}
\label{sec:BMN}
If we consider the case $\omega=0$, $\kappa =\nu $, for \eqref{eq:gensol}, this forces $\rho_0=0$
and so corresponds to the point-like BMN string rotating only in the S$^5$ 
(see \figref{fig:Quantum_spinning_string} (b)). 
As mentioned in the introduction,
this solution plays a fundamental role in our understanding the
quantum string. Here we merely calculate the one-loop correction to its classical AdS energy 
$E_0=J =\sqrt{\lambda}\kappa$. 

It is convenient to switch to Cartesian coordinates: $(\rho, \theta,\phi_1, \phi_2)\rightarrow z_k$, $k=1,...,4$ and 
$(\gamma, \psi, \varphi_1,\varphi_3)\rightarrow y_s$, $s=1,...,4$ such that
\<
\label{eq:cart_coor}
ds^2=-\frac{(1+\quarter z^2)^2}{(1-\quarter z^2)^2}dt^2+\frac{dz_kdz_k}{(1-\quarter z^2)^2}
+\frac{(1-\quarter y^2)^2}{(1+\quarter y^2)^2}d\varphi_3^2+\frac{dy_sdy_s}{(1+\quarter y^2)^2}~.
\>
Now,  expanding near $z_k=y_s=0$,
\<
t=\nu \tau+\frac{\tilde t}{\lambda^{1/4}}~,\qquad z_k=\frac{{\tilde z}_k}{\lambda^{1/4}}~, \qquad 
\varphi_2=\nu \tau+\frac{{\tilde \varphi}}{\lambda^{1/4}}~,\qquad y_s=\frac{{\tilde y}_s}{\lambda^{1/4}}~,
\>
the bosonic terms of the action \eqref{eq:action}, in conformal gauge, give the quadratic term
\footnote{We note that this is essentially the same action as that found by expanding
the action for a string in the plane-wave geometry,\cite{Metsaev:2001bj, Metsaev:2002re}, $ds^2=dx^+dx^-+\quarter x^2 dx^+dx^++dx^idx^i$
about the solution $x^+=2\nu \tau$ \cite{Berenstein:2002jq, Metsaev:2001bj, Metsaev:2002re}.}
\<
I_B=-\frac{1}{4\pi} \int d^2\sigma\, \Big[-\partial_a{\tilde t}\partial^a{\tilde t}
+\partial{\tilde \varphi}\partial^a {\tilde \varphi}+\nu^2({\tilde z}^2+{\tilde y}^2)
+\partial_a{\tilde z}_k\partial^a{\tilde z}_k+\partial_a{\tilde y}_s\partial^a{\tilde y}_s \Big]~.
\>
This action corresponds to two massless longitudinal fluctuations
${\tilde t}$ and $\tilde \varphi$, plus eight free, massive scalars, with mass $m=\nu$. 
For the fermions we find for the induced Dirac matrices $
\ind_0=\kappa~ \Gamma^-$  and $\ind_1=0$
so that the action becomes
\<
I_F=\frac{i \nu}{2\pi} \int d^2\sigma\, \Big[{\bar \theta}^1\Gamma^-\partial_+\theta^1
+{\bar \theta}^2\Gamma^-\partial_-\theta^2-2\nu {\bar \theta}^1\Gamma^-\Pi\theta^2\Big]
\>
where we have defined $\partial_\pm=\partial_0\pm\partial_1$,  $\Gamma^\pm=\mp \Gamma_0+\Gamma_9$ and  $\Pi=\Gamma_{1234}$. 
Furthermore because of the form of the fermionic kinetic operator
it was natural to choose the kappa-gauge fixing $\Gamma^+\theta^I=0$ which simplified
the mass term. This action corresponds to eight free, massive fermionic excitations,
with $m=\pm \nu$. Finally, one must include contributions from the conformal bosonic ghosts,
however for the cases in which we are interested, as was  
 shown in \cite{Drukker:2000ep, Frolov:2002av},
  the ghost contribution is essentially trivial. Their
  only effect is to cancel the two massless longitudinal bosonic fluctuations. 
  
  As the masses of the 
  transverse bosons and physical fermions are equal one immediately sees that
  the ratio of fluctuation determinants cancels and the one-loop effective action is zero. 
  Thus the correction to the AdS energy, \eqref{eq:eff_ac}, 
  $\langle E-J\rangle=\tfrac{1}{\kappa T}\Gamma$ is zero which is exactly as expected as
  this state is BPS. As we will see later, it provides a sensible 
  vacuum about which to study fluctuation interactions. 
  %
\subsection{Spinning folded string}
\label{sec:sfs}
While for the BPS solution we find zero correction to the string energy, 
a generic spinning string solution
 spontaneously breaks supersymmetry and we expect to
find a non-trivial correction at one-loop. 
We will consider  the so-called ``semi-classical scaling''
or long-string limit of the spinning string solutions, see 
\cite{Belitsky:2006en,  Frolov:2006qe} and also
\cite{Roiban:2007ju},
\<
\label{eqn:scaling_limit}
{\S}\gg {\J}\gg1, \qquad {\rm with}\qquad \ell \equiv 
\frac{\J}{2\ {\ln{\S}}}\  .
\>
As discussed at length in \cite{Frolov:2006qe, Roiban:2007ju}, upon taking $\omega=\kappa$ the solution simplifies dramatically becoming
homogeneous so that $\rho(\sigma)=\mu\sigma$. The conformal gauge condition
becomes $\kappa=\sqrt{\mu^2+\nu^2}$ and in this limit of 
large spin, $\mu=\tfrac{1}{\pi}\ln \S$ and $\ell=\tfrac{\nu}{\mu}$. 

As $\mu$ is thus very large, by rescaling the worldsheet coordinate $\sigma$ such 
that $\rho=\sigma$, we find the string length $l=2\pi \mu$ becomes infinite. The 
folded string becomes two overlapping, infinite, open strings. One can further
expand in small $\ell$, the so called 
``slow long string limit", \cite{Frolov:2006qe, Roiban:2007ju}.
In this further limit the quantum string energy is given by 
\<
E-S=\frac{\sqrt{\lambda}}{\pi}f(\lambda)\ln S~,
\>
where $f(\lambda)$ is the universal scaling function. 
At leading order this can be checked by expanding the classical 
energy which is given by $\E_0-\S=\mu\sqrt{1+\ell^2}$. We will see this form persists
at subleading orders in the semiclassical expansion, i.e. there are no
$\ln^k S$ terms, and furthermore we can calculate the numerical 
coefficients \cite{Frolov:2002av, Frolov:2006qe, Roiban:2007dq, Roiban:2007ju}
\<
f(\sql)=1-\frac{3\ln 2}{\sql}-\frac{K}{\l}+\dots
\>
where $K$ is the Catalan constant.

To calculate these coefficients we expand about the 
homogeneous, $J=0$ solution, $\hat t=\kappa \tau$, $\hat \rho=\kappa \sigma$, 
$\hat \theta=\tfrac{\pi}{2}$,
$\hat \phi_2=\kappa \tau$,  and (following \cite{Frolov:2002av} closely, where full
details can be found) we again consider the conformal gauge
action.
\paragraph{Bosons} The bosonic action \eqref{eq:action}
to quadratic order in fluctuations
 (using coordinates \eqref{eq:global_coor} 
for the AdS$_5$ space but \eqref{eq:cart_coor} for the sphere) is 
\<
I_B&=&\kern-5pt-\frac{1}{4\pi}\int d^2\sigma ~\Big[-\cosh^2 \hat \rho (\partial \tilde t)^2+\sinh^2\hat \rho(\partial \tilde \phi_2)^2
+2\kappa \sinh\hat \rho \tilde \rho (\partial_0\t t -\partial_0 \t \phi_2)\nn\\
& &+(\partial \t\rho)^2+\sinh^2\hat \rho((\partial \t \theta)^2+ \t \theta^2 (\partial  \phi_1)^2+\kappa^2 \t \theta^2)+(\partial\t\phi_3)^2+\sum_s(\partial\t y_s)^2\Big]
\>
where e.g. $(\partial t)^2=\partial_a t \partial^a t$. 
In this expression the coefficients depend on the worldsheet coordinates, 
however by making the field redefinitions 
\<
\bar \chi= \tfrac{1}{2} \sinh 2 \hat \rho~( \t \phi_2 -  \t t)~,\qquad  & &\kern-20pt\bar \xi= -\sinh^2\hat \rho~ \t \phi_2+\cosh^2\hat \rho~ \t t~,\qquad
\bar \theta=\sinh\hat \rho~ \t \theta~,\nn\\
\bar \rho=\t \rho~,\qquad & &\kern-20pt \bar x_1 =\t \theta\cos \phi_1~,  \qquad \bar x_2=\t \theta \sin\phi_1~,
\>
this can be put in the form 
\<
I_B~=~-\frac{1}{4\pi}\int d^2\sigma \kern-8pt&\Big[ &\kern-8pt(\partial \bar \chi)^2
-(\partial \bar \xi)^2+(\partial \bar \rho)^2
+4 \kappa(\partial_1\bar \chi) \bar \xi -4 \kappa (\partial_0 \bar \chi)\bar \rho \nn\\
& &+\sum_i \left((\partial \bar x_i)^2+2\kappa^2 x_i^2\right)+(\partial\t\phi_3)^2+\sum_s(\partial\t y_s)^2\Big]~.
\>
It is now straightforward to calculate the determinant of the fluctuation operator
\[
{\rm det}~K_B=-(\partial^2)^7(\partial^2+2\kappa^2)^2(\partial+4\kappa^2)
\]
corresponding to  two scalars with mass $\sqrt{2}\kappa$, one with mass $2 \kappa$ and
seven massless scalars -- two from the AdS space, five from
the sphere.
\paragraph{Fermions}  Substituting the classical solution in the expressions
for the induced Dirac matrices we find (where the flat index $0$ 
is the homologue of $t$,  $1$ corresponds to $\rho$, and $2$ to $\phi_2$)
\<
\ind_0=\kappa ~\Gamma_0\left(\cosh\hat \rho - \sinh\hat \rho ~\Gamma_{02}\right)~,\qquad
\ind_1=\kappa~ \Gamma_1~.
\>
Using the expression for the quadratic action
\eqref{eq:action}, we again find that the dependence 
on the worldsheet coordinates can be removed by a field redefinition
\<
\theta^I=S\Psi^I, \qquad{\rm with }\qquad S={\rm exp}\left(\tfrac{\kappa \sigma}{2}\Gamma_{02}\right)~,
\>
such that the corresponding transformations of the induced Dirac matrices are
\<
\tau_0=S^{-1}\ind_0S=\kappa~ \Gamma_0~,~~~{\rm and}~~~\tau_1=S^{-1}\ind_1S=\kappa~ \Gamma_1~.
\>
Making use of the relevant terms of the spin connection, $\omega_{t}{}^{ 01 }=\sinh\rho$ and
$\omega_{\phi_2}{}^{41}=\cosh\rho\cos\theta$, one can show that the portion 
of the covariant derivative that couples to the background curvature, 
${\rm D}_a=\partial_a+\tfrac{1}{4}\omega_a^{AB}\Gamma_{AB}$, essentially becomes
trivial: $S^{-1}{\rm D}_aS=\partial_a+B_a$ where $\eta^{ab}\tau_a B_b=\epsilon^{ab}\tau_a B_b=0$.
Thus the fermionic action can be written as
\<
I_F=\frac{i \sql }{2\pi}\int d^2\sigma ~(\eta^{ab}\delta^{IJ}-\epsilon^{ab}s^{IJ})
(\bar\Psi^I\tau_a\partial_b\Psi^J
+\tfrac{1}{2}\epsilon^{JK}\bar \Psi^I\tau_a\Gamma_{01234}\tau_b\Psi^K)~.
\>
As can be seen from the form of the kinetic operator one can fix the fermionic
kappa-symmetry by imposing $\Psi^1=\Psi^2=\Psi$ resulting in the 
fermion action 
\footnote{While it is not relevant for the case at hand in general one must be careful with 
the boundary conditions imposed on the fermions which can be subtle. See \cite{Mikhaylov:2010ib} for a discussion.}
\<
I_F=\frac{i \sql }{\pi}\int d^2\sigma ~
\bar\Psi^I(\tau^a\partial_a+i M) \Psi~, ~~~~{\rm where}~~~~ M=i \kappa^2 \Gamma_{234}~.
\>
Of the eight physical fermions four have mass $\kappa$ and four
have $-\kappa$, thus
\<
{\rm det}~K_F=(\partial^2+\kappa^2)^8~.
\>
\paragraph{Energy Correction}
To determine the correction to the energy we must
evaluate the sum over momenta. As we are interested 
in the leading term in the large $\kappa$ expansion
we can treat the worldsheet, after 
rescaling by $\kappa$, as having infinite extent and
so the worldsheet momenta are continuous. In momentum 
space the one-loop effective action is (having taken into account
the conformal ghosts which cancel two massless bosons)
\<
\Gamma_1=\frac{1}{2}V_2\int \frac{d^2p}{(2\pi)^2}\Big[\ln(p^2+4)+
2\ln(p^2+2)+5\ln p^2-8\ln(p^2+1)\Big]~
\>
where we recall that two-dimensional volume is 
given by $V_2=2\pi \kappa^2 \T$. While the complete expression is finite
the individual terms are divergent so we introduce a cut-off at intermediate stages 
to perform the integration. The quadratic and logarithmic divergences  
cancel and the finite result is
\<
\langle E-S\rangle\left.\right|_{\rm one-loop}=\frac{1}{\kappa \T}\Gamma_1=-\frac{3\ln2}{\pi}\ln S
\>
which is the leading correction to the universal scaling function. 
We note that the $\ln S$ dependence arises from the fact that the effective action 
is proportional to the worldsheet volume as, in the scaling limit, we 
can completely remove $\kappa$ from the action.
This remains true at all orders. 
\paragraph{Generalisations} The two-loop calculation of the universal scaling function
was carried out in \cite{Roiban:2007dq, Roiban:2007jf, Giombi:2009gd}.
The equivalence \cite{Kruczenski:2007cy} of
the spinning folded string, in the $l\rightarrow\infty$ limit, to the null cusp
Wilson loop solution \cite{Kruczenski:2002fb} plays a key role in these calculations;
as does a form of the action with particularly simple fermions \cite{Kallosh:1998ji}. 
One can obviously include the effects of non-zero
$J$ by keeping finite $\nu$, or equivalently $\ell$, dependence. The generalised
one-loop calculation in the ``long string" limit was performed in \cite{Frolov:2006qe} 
and the two-loop analysis in \cite{Roiban:2007ju, Giombi:2010fa,Giombi:2010zi}. 
Here, it is necessary to take into account the quantum corrections to the Virasoro condition
and to the relations between solution parameters and charges as described in \secref{sec:qc}. 
Furthermore, the calculation is simplified by
using a light-cone gauge \cite{Metsaev:2000yf, Metsaev:2000yu} adapted 
to a geodesic entirely in the AdS$_5$
space. These results match those found from the 
ABA \cite{Casteill:2007ct, Gromov:2008en, Bajnok:2008it}.
These calculations thus provide vigorous checks of the two-loop finiteness of the
worldsheet theory and the underlying quantum integrability. 

\subsection{Circular spinning strings}

While the energies of spinning folded strings have provided stringent
checks of ABA the relationship is slightly complicated. It is a separate
class of solutions, rigid circular spinning strings (see \cite{chapSpinning} for a review and further references), whose energies  are most transparently
related to the strong coupling expression for the S-matrix entering the ABA. 
The simplest circular strings come in two types: the so-called $\alg{su}(2)$
circular strings moving on a S$^3\subset$ S $^5$,
\cite{Frolov:2003qc}, 
and the $\alg{sl}(2)$ circular strings lying in AdS$_3\times$S$^1$ $\subset$ AdS$_5\times$S$^5$
\cite{Arutyunov:2003za}. 

The computation of the one-loop correction to the energies of the $\alg{su}(2)$ 
\cite{Frolov:2003tu, Frolov:2004bh, Beisert:2005mq} and $\alg{sl}(2)$ 
\cite{Park:2005ji, Beisert:2005cw, 
SchaferNameki:2005is, SchaferNameki:2006gk} strings 
\footnote{An early semiclassical analysis of  circular strings in AdS was
performed in \cite{deVega:1994yz}.}
played
a key part in discovering the presence of the one-loop term
 \cite{Hernandez:2006tk}
in the phase in the strong-coupling (or ``string'') form of the Bethe
Ansatz \cite{Arutyunov:2004vx, Staudacher:2004tk, Beisert:2005fw}. 

The $(S,J)$ string solution of \cite{Arutyunov:2003za} has a spiral-like shape, with projection to
$AdS_3$ being a constant radius circle (with winding number $k$), and
projection to $S^5$ -- a big circle (with winding number $m$). The
corresponding spins are, respectively, $S$ and $J$ with the Virasoro
condition implying that $u\equiv \tfrac{S}{J} = - \tfrac{m}{k}$.  Expanding the classical energy in large
semiclassical parameters $\S$ and $\J$ with fixed $k$ and $u$ \cite{Arutyunov:2003za, Park:2005ji} 
we have
\<
&& E_0 = S + J   + \frac{ \l }{ J} e_1(u,k)  + 
 \frac{ \l^2 }{ J^3} e_3(u,k) +
 \frac{ \l^2 }{ J^5} e_5(u,k) +  ...  \ .
\la{ops} 
\>
For circular strings the 
expressions for the fluctuation frequencies are sufficiently complicated that they must
be expanded in $\J$ to be evaluated and subsequently summing over modes  becomes
slightly subtle \cite{Park:2005ji, Hernandez:2005nf, Beisert:2005mq, SchaferNameki:2005tn, Beisert:2005cw, SchaferNameki:2005is, Minahan:2005qj, SchaferNameki:2006gk}. 
The correct procedure, given in \cite{Beisert:2005cw} for the $\alg{sl}(2)$ case (see also
\cite{SchaferNameki:2006gk} for the $\alg{su}(2)$ case), gives  two types of terms
for the one-loop correction, $E_1 = E_1^{\rm even} + E_1^{\rm odd}$,
where
\<
\label{eq:circ1l}
 E_1^{\rm even} = \frac{\l }{ J^2} g_2(u,k) + \frac{ \l^2}{  J^4} g_4(u,k) +...\ , \qquad E_1^{\rm
odd}= \frac{ \l^{5/2}}{  J^5} g_5(u,k) + ...~.
\la{lps} 
\>
The absence of the $\frac{1}{ J}$ and $\frac{1}{ J^3}$ terms suggests that the two
leading $\frac{ \l}{ J}$ and $\frac{ \l^2}{ J^3}$ terms receive no quantum corrections
and their coefficients should directly match weak coupling gauge theory results. Indeed, the coefficient 
$g_2$ of the  ``even''  $\frac{1}{ J^2}$ term in \rf{lps} can be reproduced as a leading $\frac{1}{ J}$ (finite spin chain length) correction from the one-loop gauge theory Bethe Ansatz
  \cite{Beisert:2005mq, Hernandez:2005nf}. 
At the same time, the presence of the non-analytic term $\frac{ \l^{5/2}}{ J^5}$ in \rf{lps} 
 implies that a similar $\frac{1}{ J^5}$ term in the
classical energy \rf{ops} is not protected so that its coefficient
cannot be directly compared to three-loop result on the gauge theory
side which implies \cite{Beisert:2005cw} that the corresponding ``string'' Bethe Ansatz
\cite{Arutyunov:2004vx} should be modified to contain a non-trivial one-loop
correction to the phase. This phase was determined by directly matching to 
higher orders in this expansion \cite{Hernandez:2006tk, Freyhult:2006vr}.

\subsection{Finite size effects and short operators}
Semiclassical analysis can also be applied to strings of finite length
and even, to a certain degree, short strings. For the folded spinning string, \secref{sec:sfs}, 
 the large $S$ corrections to the
one-loop calculation  were analysed in \cite{Beccaria:2008tg}
and the exact one-loop expression for the fluctuation determinants was found in \cite{Beccaria:2010ry}
(for two-loop results see \cite{Giombi:2010zi}). 
The one-loop correction to the small spin or short string limit
of the string 
were calculated in \cite{Tirziu:2008fk} and the generalisation
with non-zero $J$ in \cite{Beccaria:2008dq}. Short, excited strings dual to operators in the
Konishi multiplet are particularly important in testing the conjectured exact
results for the spectrum at finite volume. The correction 
to their energies at strong coupling was calculated semiclassically, with caveats
regarding the validity of these methods in this regime, in \cite{Roiban:2009aa}.
For the circular spinning strings, in addition to the energy correction \eqref{eq:circ1l}, a careful analysis
shows the presence of exponential corrections, ${\cal O}(e^{-\J})$ 
\cite{SchaferNameki:2005is, SchaferNameki:2006gk, SchaferNameki:2006ey}. Similar exponential corrections are found for quantum corrections to  finite-sized giant-magnons calculated using algebraic curve methods (see \cite{chapCurve}).
Such corrections cannot be accounted for by modifying the phase
in the BA but rather arise from finite volume effects.
See \cite{chapTrans, chapTBA}  for reviews and references. 
%
\section{Perturbative light-cone quantization}
\label{sec:plc}
As we saw in \secref{sec:BMN}, the string action expanded about the BMN string is particularly simple 
and is exactly solvable to quadratic order in fluctuations. This string solution provides
a sensible vacuum about which to perturbatively quantize the 
AdS$_5\times$S$^5$  Green-Schwarz string
\cite{Parnachev:2002kk, Callan:2003xr, Callan:2004uv, Frolov:2006cc, Hentschel:2007xn}. 
In this context it is natural to make use of light-cone gauge, introducing 
 the coordinates and  momenta, $p_\mu=h^{0a}G_{\mu \nu}\partial_a x^\nu$,
\<
\label{eq:lc_coord}
x^+=\frac{1}{2}(t+\phi)~,~~ ~x^-=\phi-t~,
~~~ p_-=\frac{1}{2}(p_\phi-p_t)~,~~ ~p_+=p_\phi+p_t
\>
where we focus on the bosonic fields for simplicity.
The Hamiltonian density ${\cal H} = p_{\mu} \dot x^{\mu} - {\cal L}$ is given by
\begin{equation}\label{hamilton}
{\cal H} = -
\frac{h^{\tau\sigma}}{h^{\tau\tau} } (x^{\prime\mu}p_{\mu})
+\frac{1 }{2 h^{\tau\tau} } ( p_{\mu} G^{\mu\nu} p_{\nu}
+ x^{\prime \mu} G_{\mu\nu} x^{\prime \nu} ) \ ,
\end{equation}
with the notation $x'=\partial_\sigma x$ and $\dot x=\partial_\tau x$. 
As is usual in theories with general coordinate invariance, 
the Hamiltonian is a sum of constraints times Lagrange multipliers.

To impose light-cone gauge one sets $x^+ = \tau$ and
$p_-= {\rm const}$.  
The metric coefficients ${1 }/{h^{\tau\tau} }$ and
${h^{\tau\sigma}}/{ h^{\tau\tau} }$ act as Lagrange multipliers,
generating delta functions that impose two constraints
which determine $x^-$ 
and $p_+$ in
terms of the transverse variables (and the constant
$p_-$). 
\footnote{In fact, the constraints determine the 
derivatives of  $x^-$ and 
 so $x^-$ itself is non-local in this gauge. This has important 
consequences for the ``off-shell'' symmetry algebra.} 
The transverse
coordinates and momenta $x^A, ~ p_A$ $A=1,\dots,8$ will then have dynamics which follow from the
light-cone Hamiltonian
$- p_+= {\cal H}_{lc}$.
The first constraint, or level-matching constraint, yields
$x^{\prime -}=-x^{\prime \, A} p_A/p_-$. 
%
%
While solving the quadratic constraint equation
for $p_+$ we obtain the somewhat dispiriting result
\<
\label{hamiltonianLC}
-{\cal H}_{\rm lc} = \frac{p_- G_{+-}}{ G_{--} } +
\frac{p_- \sqrt{G}}{ {G_{--}} } \sqrt{ 1 + \frac{G_{--}}{p_-^2}
( p_A G^{AB} p_B + x^{\prime A} G_{AB} x^{\prime B} ) +
\frac{G_{--}^2}{p_-^4}(x^{\prime A} p_A)^2 }\ ,
\>
with 
$G \equiv G_{+-}^2 - G_{++} G_{--}$.
\footnote{We have made use of 
the fact that the \adss metric, \eqref{eq:cart_coor}, rewritten 
in light-cone coordinates, \eqref{eq:lc_coord}, has no
 $G_{+A}$ or $G_{-A}$ components.}
 Using the
 relation between the canonical momenta and the target space charges we have
 \<
 E-J=-P_+=\tfrac{\sqrt{\lambda}}{2\pi}\int_0^{2\pi} d\sigma \ {\cal H}_{\rm lc}~,\qquad 
 \tfrac{1}{2}(E+J)=P_-=\tfrac{\sqrt{\lambda}}{2\pi}\int_0^{2\pi}d\sigma\ p_- ~.
 \>

\paragraph{Perturbative expansion} To make progress we perform the large tension expansion:
 rescaling  the transverse fields by $\lambda^{-1/4}$ 
and expanding in large $\sql$, or equivalently $P_-=\sqrt{\lambda}p_-\sim J$, 
while keeping $-P_+=E-J$ fixed. Being careful with the expansion
of the $G_{--}$ terms, see e.g. \cite{Callan:2004uv}, one finds the first two orders,
\<
\label{pertHam}
{\cal H}_{\rm lc}^{pp} & = & \frac{1}{2p_-}\left[ (\dot p^A)^2
    + ({ x'}^A)^2 +p_-^2 (x^A)^2 \right]\nn\\
    & &+
\frac{1}{4\sqrt{\lambda} p_-}\left( z^{2} (p_y^{2}+y'^2)-y^{2}  (p_z^{2}+z'^2)+
2 z^2{z}'^2  - 2 y^2 {y}'^2\right)~,
\>
where beyond leading order the eight transverse fields split into two sets of four, $x^A=(z^i,y^s)$. 
One can remove the dependence on the density $p_-$ by rescaling the 
worldsheet coordinates, and thus we see that we are taking the large
charge limit but keeping the worldsheet compact. 

The leading order
term is simply the plane-wave Hamiltonian whose spectrum consists of 
an infinite tower of non-interacting massive oscillators,  
\<
x^A(\sigma,\tau) = 
    \sum_{n=-\infty}^\infty x_n^A(\tau) e^{-i n \sigma}~,\qquad
x_n^A(\tau) =  \frac{i}{\sqrt{2 \omega_n}} (a_n^A e^{-i \omega_n \tau}
    - {a_{-n}^{A\dagger}} e^{i \omega_n \tau} )\ ,
\>
where $n\in \mathbb{Z}$, $\omega_n = \sqrt{ p_-^2 + n^2}$,
and the raising and lowering operators obey the usual commutation relations.
One can straightforwardly include the fermions, though 
the subleading interaction terms become somewhat involved
\cite{Callan:2003xr, Callan:2004uv, Frolov:2006cc}. At leading order one 
again gets massive oscillators, $b^\alpha_n$,
$\alpha=1,\dots, 8$ and thus
the full plane-wave Hamiltonian, ${H}_{\rm pp}$, is
\begin{eqnarray}
{H}_{\rm pp} & = & \frac{1}{p_-} \sum_{n=-\infty}^\infty \omega_n
   \left( {a_n^A}^\dagger a_n^A +  {b_n^\alpha}^\dagger b_n^\alpha \right)\ ,
\end{eqnarray}
where one can immediately see that the energy of 
the vacuum state, $|{\rm Vac}\rangle$, corresponding to a string with
charge $P_-$ vanishes.
\paragraph{Near-BMN energy spectrum} 
The quartic terms give rise to corrections of order ${\cal O}(1/J)$,
the effects of which can
be  perturbatively included in the spectrum. In the simple case
where we consider a single complex boson from the sphere
$y=y^1+i y^2$, the leading correction to the two excitation state 
$a_n^\dagger a_{-n}^\dagger | P_-\rangle $ is 
\<
E-J=2 \sqrt{1+\l' n^2}-2\frac{\l' n^2}{J}+\frac{N_B(n^2)}{J}
\>
with $\l'=\l/J^2$ an effective coupling. Due to the form of the interactions 
there is a normal ordering ambiguity, here characterised by the arbitrary function
$N_B(n^2)$. There are related functions in the correction 
to all energies and they are fixed by demanding that the full spectrum possess the underlying
global $\alg{psu}(2,2|4)$ symmetry. This implies, for example, $N_B=0$. 
Equivalently, they could be fixed by demanding that
the algebra of generators, including the Hamiltonian, is satisfied at this order. 
These expressions for string energies can be compared to the string ABA \cite{Callan:2004ev, McLoughlin:2004dh, Arutyunov:2005hd, Frolov:2006cc, Hentschel:2007xn} and were one of the first pieces of evidence for a non-trivial dressing phase interpolating between strong and weak coupling. 
%
\subsection{Worldsheet S-matrix}

As the theory in  light-cone gauge has
only massive particles, we can study the interactions by calculating the
worldsheet S-matrix. Modulo issues of gauge dependence 
\footnote{The
S-matrix is gauge-dependent, since unlike the spectrum it is not a
physical object with a clear target-space interpretation. The differences
between gauges can be attributed to the definition of
the string length \cite{Staudacher:2004tk}. 
The difference in the definition of length and the gauge-dependence of the S-matrix,
mutually cancel in the Bethe equations
\cite{Frolov:2006cc, Arutyunov:2006iu}.}
this object should match the spin chain S-matrix introduced in 
\cite{Staudacher:2004tk}, see \cite{chapSMat, chapSProp} for reviews. 
The perturbative study of the worldsheet S-matrix was initiated in 
\cite{Klose:2006zd} 
while  its symmetries 
and many properties were analysed in \cite{Arutyunov:2006ak, Arutyunov:2006yd} (see \cite{Arutyunov:2009ga} for an extensive review).  To define the S-matrix one must
consider the theory on the plane: this corresponds to scaling
 $p_-$ out of  the action and taking the decompactification limit
 $p_-\rightarrow\infty$. In order to define free, asymptotic states for generic momentum
 one relaxes the level matching condition and then studies the 
 interactions in powers
of $\sql$ or equivalently in a small (worldsheet) momentum 
 expansion. 
 
\paragraph{Asymptotic states}
%
%
Of the global group, the light-cone gauge preserves a subset $
\grPSU(2|2)_L\times \grPSU(2|2)_R\subset \grPSU(2,2|4 )$. The bosonic subgroup of each
$\grPSU(2|2)$ factor consists of two $\grSU(2)$ groups and 
it is useful to introduce a bispinor notation for the physical bosons  $
 Z_{\laa\raa} = (\sigma_i)_{\laa\raa} z^i$ , $Y_{\lAA\rAA} = (\sigma_s)_{\lAA\rAA} y^s$
 and fermions,  $\Psi_{\lAA\raa}, \Bsi_{\laa\rAA}$, which are charged under different 
 combinations of the $\grSU(2)$'s.
One may define superindices $A=(\lAA|\laa)$ and ${\dot
A}=(\rAA|\raa)$ combining
all asymptotic  fields creating incoming or outgoing particles
into a single bi-fundamental supermultiplet of
which we will denote by $\Phi^{(in/out)}_{A{\dot A}}$.
%

%
%

\paragraph{The S-matrix.} The two-particle S-matrix is a unitary operator
relating $in$- and $out$-states. 
In the basis  $\Phi_{A{\dot A}}(p)$, so that
$\ket{\Phi_{A\dot{A}}(p) \Phi_{B\dot{B}}(p')}^{(in)}=\Phi^{(in)}_{A\dot{A}}(p) \Phi^{(in)}_{B\dot{B}}(p') \ket{{\rm Vac}}$,
 its matrix representation is
\[\label{smatrix_components}
 \Smatrix \, \ket{\Phi_{A\dot{A}}(p) \Phi_{B\dot{B}}(p')}^{(in)}
 = \ket{\Phi_{C\dot{C}}(p) \Phi_{D\dot{D}}(p')}^{(out)} \,
\Smatrix_{A\dot{A}B\dot{B}}^{C\dot{C}D\dot{D}}(p,p')~.
\]
Before gauge fixing the worldsheet theory is classically
integrable \cite{Bena:2003wd}; since fixing light-cone may be interpreted as expanding about
the BMN solution and solving some of the equations of motion, the gauge-fixed theory
is also expected to be integrable at the classical level.
In such an integrable theory, the S-matrix, invariant under a non-simple product
group, must be a tensor
product of S-matrices for each of the factors (see e.g. \cite{Ogievetsky:1987vv})%
\footnote{This can be understood as a requirement that the
Faddeev-Zamolodchikov algebra is also a direct product: the field
$\Phi_{A{\dot A}}$ is represented by a bilinear in oscillators:
$\Phi_{A{\dot A}}\sim z_Az_{\dot{A}}$ each transforming under one of
the $\grPSU(2|2)$ factors \cite{Arutyunov:2006yd}. The two sets of oscillators mutually
commute. The braiding relations for each of these sets are
determined by an $\grPSU(2|2)$-invariant S-matrix $\smatrix$
consistent with the Lagrangian of the theory.}
\<
\Smatrix=\smatrix\otimes \smatrix \comma
\Smatrix_{A\dot{A}B\dot{B}}^{C\dot{C}D\dot{D}}(p,p')=
\smatrix_{AB}^{CD}(p,p')\smatrix_{\dot{A}\dot{B}}^{\dot{C}\dot{D}}(p,p')
\; .
\label{factorise_0}
\>
It is important to note that a factorised tensor structure does not
follow solely from the $\grPSU(2|2)\times\grPSU(2|2)$ symmetry
considerations; confirming group factorisation is thus an
important test of integrability.

The first nontrivial order in the expansion of the S-matrix in the
coupling constant $2\pi/\sqrt{\lambda}$ defines the
T-matrix
\[ \label{asmatrix_expansion}
 \Smatrix = \unit + \frac{2\pi i}{\sqrt{\lambda }} \, \Tmatrix +
\order\lrbrk{ \frac{1}{\lambda} } \; .
\]
which inherits the factorised form $\Tmatrix = \unit\otimes\tmatrix + \tmatrix\otimes\unit$
from the S-matrix.
Furthermore, since $\grSU(2)\times\grSU(2)\subset\grPSU(2|2)$
is a manifest symmetry of the gauge-fixed worldsheet theory,
$\tmatrix$ may be parametrised in terms of ten unknown functions
of the momenta $p$ and $p'$. 
These functions, to leading order in $1/\sqrt{\lambda}$, 
 can be easily extracted 
 from  the matrix elements of 
 quartic terms  of the light-cone Hamiltonian \eqref{pertHam} (see \cite{Klose:2006zd}
 where explicit expressions for $\tmatrix$ can be found). Equivalently one can 
 Legendre transform with respect to the transverse fields to find the light-cone
 Lagrangian and then use the usual LSZ reduction to calculate the worldsheet 
 scattering amplitudes perturbatively.  

\paragraph{Properties of the S-matrix}
\begin{itemize}
\item The explicit perturbative calculation does indeed show that 
the two-body S-matrix has the factorised form \eqref{factorise_0}. Furthermore, it can 
be explicitly checked to leading order that the ten functions in the T-matrix
agree with the corresponding functions in the strong coupling BA S-matrix.
It can be shown explicity that there is no two-to-four particle scattering \cite{Klose:2006zd}.

\item In calculating the S-matrix we relax the level-matching constraint. In 
this ``off-shell" formulation of the theory the symmetries become extended by two 
additional central charges related to the worldsheet momentum \cite{Arutyunov:2006ak} 
(the same as found in the spin chain \cite{Beisert:2005tm}).
Furthermore, as the supersymmetry generators, $Q\sim \int e^{i x^-}\Omega(Z,Y, \Bsi, \Psi)$,   
depend on the zero mode of the longitudinal coordinate, $x^-\sim\int d\sigma \partial_\sigma x^-$, there is a mild non-locality in the action of the symmetries which thus satisfy a Hopf algebra 
\cite{Klose:2006zd, Arutyunov:2006yd}.

\item The integrable structures of the perturbative string S-matrix have been further studied
including the construction of the classical r-matrix e.g. \cite{Beisert:2007ty}. Furthermore, 
assuming the quantum integrability of the full worldsheet theory, and using the 
global symmetries, the worldsheet S-matrix  
was uniquely determined up to an overall phase.
We refer the reader to \cite{chapSMat, chapSProp, chapYang} for a more complete discussion  of
these and other exact properties of the worldsheet S-matrix. 

\end{itemize}

\subsection{Simplifying Limits}

Due to the complexity of the world sheet theory, going beyond the leading
perturbative term is challenging. One simplifying limit which has proved useful is the
``near-flat limit'' \cite{Maldacena:2006rv}. This limit corresponds to studying the
worldsheet near a constant density solution boosted with rapidity $\lambda^{1/4}$ 
in the worldsheet light-cone direction, $\sigma^-$. The left- and right-moving excitations on the 
worldsheet scale differently and the right movers essentially decouple. 
The resulting theory has only quartic interactions and is much more tractable. 
The one-loop and two-loop \cite{Klose:2007wq, Klose:2007rz, Puletti:2007hq}
corrections to the S-matrix have been calculated and shown to match the all-order
conjecture \cite{Beisert:2006ib}; furthermore factorization at one-loop was explicitly shown.
In the two-loop calculation radiative corrections induce a correction to the relativistic dispersion 
 relation which corresponds to the expansion of the sine function, natural from a
 spin chain perspective, which appears
 in the exact dispersion relation \cite{Beisert:2005tm}. 
 
 Another interesting formulation of the theory is found 
 via a generalisation of the Pohlmeyer reduction 
 \cite{Pohlmeyer:1975nb} which is used to relate, at a classical level, the string theory 
 on \adss to a massive,  Lorentz invariant theory which only involves the physical 
 fields. Applied to strings on 
 $\mathbb{R} \times S^3$ this method consists of gauge fixing and solving the Virasoro constraints 
 so that the remaining degree of freedom satisfies the sine-Gordon equation of motion 
 \cite{Tseytlin:2003ii, Mikhailov:2005qv}. Generalised to the full superstring
 \cite{Grigoriev:2007bu, 
 Mikhailov:2007xr, Grigoriev:2008jq, Miramontes:2008wt} the reduced theory is a massive deformation
 of a gauged WZW model with an integrable potential. 
 The resulting model has been explicitly shown to be UV finite
 to two-loops and there is evidence that 
  the equivalence to the standard formulation persists at the 
  quantum level \cite{Roiban:2009vh, Hoare:2009rq}.
The two-particle S-matrix was calculated in this formalism in \cite{Hoare:2009fs, Hoare:2010fb} where it was shown that   
it has the appropriate group factorisation properties. Being manifestly
 Lorentz invariant this formalism may provide a better basis for understanding the quantum theory. 
 
\vspace{0.5cm} \noindent {\bf Acknowledgements:} It is a pleasure to thank 
N.\ Beisert, T.\ Klose, M.\ Magro and A.\ Tseytlin for useful comments on the manuscript. 

\phantomsection
\addcontentsline{toc}{section}{\refname}
\bibliography{intads,chapters}
\bibliographystyle{nb}

\end{document}